\documentclass[letter,bibyear]{aa} 

\usepackage{natbib}
\usepackage{graphicx}
\usepackage{txfonts}

\newcommand{\cpl}{Chem. Phys. Lett.}

\newcommand{\jms}{J.~Mol.~Spectrosc.}

\newcommand{\jmst}{J.~Mol.~Struct.}

\newcommand{\kms}{km s$^{-1}$}

\begin{document}

\title{QUIJOTE discovery of the cation radicals HC$_5$N$^+$ and HC$_7$N$^+$\thanks{Based
on observations with the Yebes 40m telescope (projects 19A003, 20A014, 20D023, 21A011, 21D005,
22A007, 22B029, and 23A024). The 40 m radio telescope at Yebes Observatory is operated by the Spanish Geographic Institute (IGN; Ministerio de Transportes y Movilidad Sostenible).
 }}

\author{
J.~Cernicharo\inst{1},
C.~Cabezas\inst{1},
M.~Ag\'undez\inst{1},
Y.~Endo\inst{2},
B.~Tercero\inst{3,4},
N.~Marcelino\inst{3,4}, and
P.~de~Vicente\inst{4}
}

\institute{Dept. de Astrof\'isica Molecular, Instituto de F\'isica Fundamental (IFF-CSIC),
C/ Serrano 121, 28006 Madrid, Spain. \newline \email jose.cernicharo@csic.es
\and Department of Applied Chemistry, Science Building II, National Yang Ming Chiao Tung University, 1001 Ta-Hsueh Rd., Hsinchu 300098, Taiwan
\and Observatorio Astron\'omico Nacional (OAN, IGN), C/ Alfonso XII, 3, 28014, Madrid, Spain.
\and Observatorio de Yebes (IGN). Cerro de la Palera s/n, 19141 Yebes, Guadalajara, Spain
}

\date{Received: May 20, 2024; accepted: May 31, 2024}

\abstract{We present the discovery with the QUIJOTE line survey of the cations HC$_5$N$^+$
and HC$_7$N$^+$ in the direction of TMC-1. Seven lines with half-integer quantum numbers from 
$J$=25/2-23/2 to 37/2-35/2 have been assigned to HC$_5$N$^+$ and eight lines from $J$=55/2-53/2 to 71/2-69/2 
to HC$_7$N$^+$.
Both species have inverted $^2\Pi$ ground electronic states with very good estimates for their $B_0$
and $A_{SO}$ constants based on optical observations.
The
lines with the lowest $J$ of HC$_5$N$^+$ exhibit multiple components due to the
hyperfine structure introduced by the H and N nuclei. However, these different components collapse
for the higher $J$. No hyperfine structure is found for any of the lines of HC$_7$N$^+$.
The derived effective rotational and distortion constants for HC$_5$N$^+$ are
$B_{\rm eff}$\,=\,1336.662\,$\pm$\,0.001 MHz
and $D_{\rm eff}$\,=\,27.4\,$\pm$\,2.6 Hz, while for HC$_7$N$^+$ they are $B_{\rm eff}$\,=\,567.85036\,$\pm$\,0.00037 MHz and
$D_{\rm eff}$\,=\,4.01\,$\pm$\,0.19 Hz. From the observed intensities, we derived
$T_{\rm rot}$\,=\,5.5\,$\pm$\,0.5\,K and  $N$\,=\,(9.9\,$\pm$\,1.0)\,$\times$\,10$^{10}$ cm$^{-2}$ for HC$_5$N$^+$, while 
we obtained
$T_{\rm rot}$\,=\,8.5\,$\pm$\,0.5\,K and $N$\,=\,(2.3\,$\pm$\,0.2)\,$\times$\,10$^{10}$ cm$^{-2}$ for HC$_7$N$^+$.
The HC$_5$N/HC$_5$N$^+$, C$_5$N/HC$_5$N$^+$, C$_5$N$^-$/HC$_5$N$^+$, HC$_7$N/HC$_7$N$^+$, HC$_5$N$^+$/HC$_7$N$^+$, and C$_7$N$^-$/HC$_7$N$^+$
abundance ratios are 670\,$\pm$\,80, 4.8\,$\pm$\,0.8, 1.2\,$\pm$\,0.2, 1000\,$\pm$\,150, 4.2\,$\pm$\,0.5, and 2.2\,$\pm$\,0.2,
respectively. We have run chemical modelling calculations to investigate the formation and destruction of these new cations. 
We find that these species are mainly formed through the reactions of H$_2$ and the cations C$_5$N$^+$ and C$_7$N$^+$, 
and by the reactions of H$^+$ with HC$_5$N and HC$_7$N, while they are mostly destroyed through a reaction with H$_2$ 
and a dissociative recombination with electrons. Based on the underestimation of the abundances of HC$_5$N$^+$ and 
HC$_7$N$^+$ by the chemical model by a factor $\sim$\,20, we suggest that the rate coefficients currently assumed 
for the reactions of these cations with H$_2$ could be too high by the same factor, something that will be worth 
investigating.}


\titlerunning{Cyanodiacetylene cation}
\authorrunning{Cernicharo et al.}

\maketitle

\section{Introduction}
Among the new species discovered in the last years with the QUIJOTE\footnote{\textbf{Q}-band \textbf{U}ltrasensitive \textbf{I}nspection \textbf{J}ourney
to the \textbf{O}bscure \textbf{T}MC-1 \textbf{E}nvironment} line survey of TMC-1
\citep[][and references therein]{Cernicharo2021a,Cernicharo2023a,Cernicharo2023b},
it is worth noting that ten of them are the protonated form of abundant species. These are HC$_5$NH$^+$ \citep[protonated HC$_5$N,][]{Marcelino2020}, HC$_3$O$^+$ \citep[protonated C$_3$O,][]{Cernicharo2020a}, HCCNCH$^+$ 
\citep[protonated HCCNC,][]{Agundez2022}, HCCS$^+$ \citep[protonated CCS,][]{Cabezas2022a}, HC$_3$S$^+$ \citep[protonated CCCS,][]{Cernicharo2021b}, CH$_3$CO$^+$ \citep[protonated H$_2$CCO,][]{Cernicharo2021c}, HC$_7$NH$^+$
\citep[protonated HC$_7$N,][]{Cabezas2022b}, NC$_4$NH$^+$ \citep[protonated form of the non-polar molecule NC$_4$N,][]{Agundez2023a}, C$_5$H$^+$ \citep[protonated form of non-polar C$_5$,][]{Cernicharo2022a}, and H$_2$C$_3$H$^+$ \citep[protonated $l$-H$_2$C$_3$,][]{Silva2023}.
The observational and theoretical status of protonated molecules in cold dense clouds indicate that protonated-to-neutral abundance ratios MH$^+$/M are in the range 10$^{-3}$-10$^{-1}$ for neutral molecules M with proton affinities above that of CO, and lower ratios for molecules M with lower proton affinities \citep{Agundez2022}. It has also been found that chemical models tend to underestimate MH$^+$/M ratios, which suggest the existence of alternative paths to direct protonation for the formation of these species MH$^+$.

The abundant radicals C$_n$H and C$_n$N are species with a high electron affinity, leading to the 
formation -- through electron attachment -- of the anionic species C$_4$H$^-$ \citep{Cernicharo2007}, 
C$_6$H$^-$ \citep{McCarthy2006}, C$_8$H$^-$ \citep{Brunken2007,Kawaguchi2007,Remijan2007}, 
C$_{10}$H$^-$ \citep{Remijan2023,Pardo2023}, CN$^-$ \citep{Agundez2010}, C$_3$N$^-$ \citep{Thaddeus2008}, 
C$_5$N$^-$ \citep{Cernicharo2008,Cernicharo2020b}, and C$_7$N$^-$ \citep{Cernicharo2023a}. The same radicals 
could also produce, 
via protonation, the HC$_n$H$^+$ and HC$_n$N$^+$ radical cations. These species have escaped detection due 
to the lack of a dipole moment for the HC$_n$H$^+$ symmetric species, and the lack of rotational spectroscopy 
for the HC$_n$N$^+$ ones. These species could play an important role in the chemistry of interstellar clouds 
and the detection of the polar HC$_n$N$^+$ species could be achieved with sensitive line surveys such as QUIJOTE.

In this Letter we present the discovery -- for the first time in space -- of the
cationic radicals HC$_5$N$^+$ and HC$_7$N$^+$ through the observation of seven and eight of their rotational transitions, respectively.
We discuss the possible formation and destruction paths of the cations and conclude that the 
adopted reaction rates
of H$_2$ with  these cationic radicals limit their modelled abundances. Other possible routes
for their formation, such as the reaction of CN
with the cations C$_{2n}$H$_2$$^+$, have also been explored.

\section{Observations}
The observational data used in this work are part of QUIJOTE \citep{Cernicharo2021a},
a spectral line survey of TMC-1 in the Q band carried out with the Yebes 40m telescope at
the position $\alpha_{J2000}=4^{\rm h} 41^{\rm  m} 41.9^{\rm s}$ and $\delta_{J2000}=
+25^\circ 41' 27.0''$, corresponding to the cyanopolyyne peak (CP) in TMC-1.
The receiver
was built within the Nanocosmos project\footnote{\texttt{https://nanocosmos.iff.csic.es/}}
and consists of two cold high-electron mobility transistor amplifiers covering the
31.0-50.3 GHz band with horizontal and vertical polarizations.
Receiver temperatures
vary between 16\,K at 32 GHz and 30\,K at 50 GHz. The back ends are
$2\times8\times2.5$ GHz fast Fourier transform
spectrometers with a spectral resolution of 38 kHz, providing the whole coverage
of the Q band in both polarizations.
A detailed description of the system
is given by \citet{Tercero2021}, and details on the QUIJOTE line survey observations
have been previously provided \citep{Cernicharo2021a,Cernicharo2023a,Cernicharo2023b,Cernicharo2024a,
Cernicharo2024b}. The frequency switching method has been used for all observations.
The data analysis procedure has been described by \citet{Cernicharo2022b}.
The total observing time on source is 1202 hours and the
measured sensitivity varies between 0.07 mK at 32 GHz and 0.2 mK at 49.5 GHz.

The main beam efficiency can be given across the Q band by
$\eta_{\rm eff}$=0.797 exp[$-$($\nu$(GHz)/71.1)$^2$]. The
forward telescope efficiency is 0.95 and
the beam size at half power intensity is 54.4$''$ and 36.4$''$
at 32.4 and 48.4 GHz, respectively.

The absolute intensity calibration uncertainty is 10$\%$. However, the relative
calibration between lines within the QUIJOTE survey is certainly better because all
of them are observed simultaneously and have the same calibration uncertainties and systematic
effects.
The data were analysed with the GILDAS package\footnote{\texttt{http://www.iram.fr/IRAMFR/GILDAS}}.

\begin{figure}
\centering
\includegraphics[width=\columnwidth]{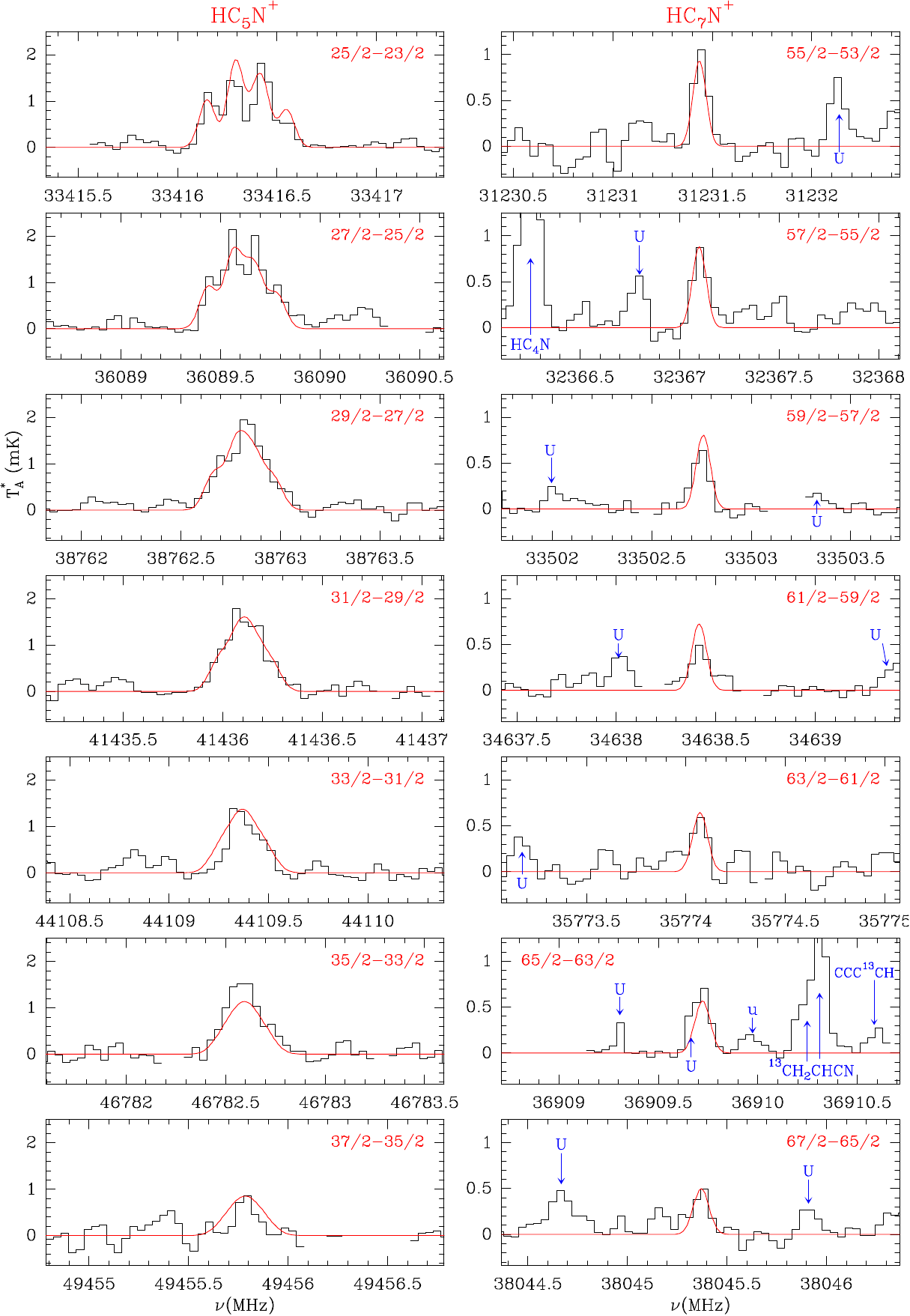}
\caption{Observed transitions of HC$_5$N$^+$ (left column) and HC$_7$N$^+$ (right column)
in TMC-1. Quantum numbers are indicated at the
top right of each panel.
The abscissa corresponds to the rest frequency. The ordinate is the antenna temperature, corrected for
atmospheric and telescope losses, in 
milli Kelvin.
Blanked channels correspond to negative features produced when folding the frequency-switched data.
The red line shows the computed synthetic spectra for the lines of the two species
(see Sect. \ref{sec:results_hc5n+} and \ref{sec:results_hc7n+}). The physical parameters used for the
models are given in Sect. \ref{column_densities}. The centroids of the modelled lines of HC$_5$N$^+$ have
been fixed to the observed ones (see Table \ref{line_centroid}) and the simulated hyperfine components
have been added to these centroids. For HC$_7$N$^+$ the line frequencies are given in Table \ref{line_parameters_hc7n+}.}
\label{fig_hc5n+}
\end{figure}

\section{Results}
Line identification in this work has been performed using the 
MADEX code \citep{Cernicharo2012b}, in addition to the CDMS
and JPL catalogues \citep{Muller2005,Pickett1998}.
The intensity scale
utilized in this study is the antenna temperature ($T_A^*$). Consequently, the telescope parameters and source
properties have been
used when modelling the emission of the different species to produce synthetic spectra in this
temperature scale. In this work we assume
a velocity for the source relative to the
local standard of rest of 5.83 \kms\, \citep{Cernicharo2020c}. The source is assumed to be circular
with a uniform brightness temperature and a radius of 40$''$ \citep{Fosse2001}.
The procedure to derive line parameters is described in Appendix\,\ref{app:lineparameters}.
The observed line intensities have been modelled using a local thermodynamical equilibrium
(LTE) hypothesis, or a large
velocity gradient (LVG) approach. In the later case, MADEX uses the formalism described by \citet{Goldreich1974}.

\subsection{Discovery of HC$_5$N$^+$}\label{sec:results_hc5n+}

Among the strongest unknown features in the QUIJOTE line survey, we have found a series of
seven lines in harmonic relation with half-integer quantum numbers from $J_u$=25/2 to 37/2 
(see Fig. \ref{fig_hc5n+}; $J_u$ and $J_l$ denote the upper and lower
quantum numbers, respectively). The first two transitions ($J_u$=25/2 and 27/2) of this molecule show four 
fine and/or hyperfine structure components, while the rest of the transitions show a remarkable line 
broadening. The frequency centroids of these lines have been derived and are given in Table \ref{line_centroid}. 
Taking into account the considerable variation of line profiles among the transitions, we estimate an 
uncertainty of 40 kHz for these frequency centroids. They can be fitted with the standard formula
$\nu$=2\,$B_{\rm eff}$\,$J$\,$-$\,4\,$D_{\rm eff}$\,$J^3$, with
$B_{eff}$=$B_0$(1 $\pm$ $B_0$/$A_{SO}$)  \citep{Townes1975}. We derived
$B_{\rm eff}$=1336.662$\pm$0.001 MHz and $D_{\rm eff}$=27.4$\pm$2.6 Hz.
These parameters have to be considered as effective as they reproduce the centroid of the line, which can be affected by the fine and hyperfine structure of the observed transitions. The standard deviation of the fit is 26 kHz. The total integrated intensity, mean velocity, and equivalent line width of the observed lines are given in Table \ref{line_centroid}.
We have searched for other lines with a similar line profile over the entire QUIJOTE frequency coverage and none have been found. We have also checked that a molecule with $B$\,=\,$B_{\rm eff}$/2 and integer quantum numbers cannot be assigned to our data because 
all lines with $J$ being even
would be missing. Given the excellent fit obtained just with two parameters, we can conclude that the observed transitions are due to a linear open shell molecule
with a $^2\Pi$ electronic ground state and $B\sim$1336.7 MHz (hereafter, referred to as B1336)
The lines are not detected towards the carbon-rich evolved star IRC+10216, a source rich in
cyanopolyynes molecules (HC$_{2n+1}$N), C$_n$H and C$_n$N radicals, and their
anions \citep[see, e.g.][and references therein]{Cernicharo2020b,Cernicharo2023a,Remijan2023,Pardo2023}.

The derived rotational constant is 0.3\% larger than that of HC$_5$N, while the distortion constant is very close to that of this species 
($B$=1331.333 MHz, $D$=30.1 Hz; \citealt{Bizzocchi2004}). The excellent agreement with the distortion constant of HC$_5$N 
and that of other similar species (see App. \ref{candidates_hc5n+})
gives strong support to a linear or quasi-linear open shell species. In Appendix \ref{candidates_hc5n+} we discuss all 
possible candidates for the series of B1336 lines.
The most promising candidate is the cationic radical HC$_5$N$^+$. Unfortunately, its rotational spectrum has not been observed in the laboratory. However, it has been observed in the optical through its electronic $X^2\Pi$$\rightarrow$$A^2\Pi$ transition by \citet{Sinclair1999a}. Their spectral resolution allowed for the rotational structure of the electronic band providing a rotational constant $B_0$\,=\,1337.5\,$\pm$\,0.2 MHz to be resolved, in excellent agreement with that derived for B1336. They also derived the spin-orbit constant, $A_{SO}$, to be $-$35.71\,$\pm$\,0.41 cm$^{-1}$ but no $\Lambda$ doubling was observed. Hence, HC$_5$N$^+$ has an inverted $^2\Pi$ 
electronic ground state, that is,
the $\Omega$=3/2 ladder is the lowest in energy, with the $\Omega$=1/2 one more than 50 K above it. Hence, the lines from 
the $^2\Pi_{1/2}$ ladder are 
extremely weak in TMC-1 due to the low kinetic temperature of this cloud
\citep[$\sim$9\,K,][]{Agundez2023b}. 

On this basis, we conclude that the lines observed in TMC-1 correspond to the rotational transitions of HC$_5$N$^+$ in its $^2\Pi_{3/2}$ ladder. Taking into account the $A_{SO}$ constant derived by \citet{Sinclair1999a}, the rotational constant $B_0$ of 
HC$_5$N$^+$ in its ground vibrational state can be derived from the relation $B_{\rm eff}$\,=\,$B_0$\,(1\,+\,$B_0$/$A_{SO}$) to be 
$B_0$\,=\,1338.331$\pm$0.001 MHz, which is in very good agreement with the value derived by these authors from less accurate optical observations. Quantum chemical calculations for HC$_5$N$^+$ indicate that the molecule has a large dipole moment between 6.5 and 6.9\,D (see Appendix\,\ref{ab_initio}, \citealt{Zhang2012}, and \citealt{Gans2019}).

The lack of nearby features similar to those observed indicates that the $\Lambda$-doubling, if any, 
is small (see Fig. \ref{fig_hc5n+}). In fact, we can discard a significant fine structure as the 
first electronic $^2\Sigma$ state has an energy above $X^2\Pi$ of $\sim$21800 cm$^{-1}$  \citep{Gans2019}. 
The $\Lambda$-doubling for a transition $J$+1$\rightarrow$$J$ of a $^2\Pi_{3/2}$ state produces a 
splitting between the fine structure components that can be approached as 
$\pm$3$q$$B$/$A_{SO}$($J$+1/2)($J$+3/2), with $q$\,=\,2$B^2$/$\Delta$($^2\Sigma-^2\Pi$), where $\Delta$($^2\Sigma-^2\Pi$) is 
the energy difference between the $^2\Sigma$ and the $^2\Pi$ electronic states \citep{Mulliken1931}. Using the values 
given above, we can compute the expected splitting for the $J$=37/2-35/2 transition to be $\pm$8 kHz, that is, much 
smaller than our spectral resolution. Hence, the spectral pattern observed for the lowest frequency transitions of 
our series must correspond to the 
magnetic hyperfine effects due to the interaction between H and N nuclear spin with the
electron orbital and electron spin angular momenta. The hyperfine splitting produces several lines, many 
of them mutually blended due to the high $J$ values of the observed transitions.

Although the identification of HC$_5$N$^+$ appears robust, the assignment of the quantum numbers to the hyperfine lines 
observed in TMC-1 is rather difficult. Hence, we have simulated the expected spectrum with the SPFIT code 
\citep{Pickett1991} using the results of our quantum chemical calculations for the hyperfine constants as 
described in Appendix\,\ref{ab_initio}. The rotational and distortion constants have been fixed to 
the values derived from the frequency centroids (see above). Fig. \ref{simul} shows the simulated 
spectra together with the observations for the first four transitions. Although the agreement between 
the synthetic spectrum and observations is not perfect, a similarity between the patterns can be observed. 
Slightly different values for H and N hyperfine parameters from those predicted by quantum chemical 
calculations reproduce the observed patterns and line shapes satisfactorily, as can be seen in 
Fig. \ref{fig_hc5n+} and \ref{simul}. The estimated hyperfine parameters are given in Table \ref{hyperfine}. 
Although the agreement between the modelled and observed lines is good and fully supports the identification 
of B1336 with HC$_5$N$^+$, only laboratory observations of the 
low-$J$ rotational transitions of this radical will 
permit accurate values of the hyperfine constants to be determined.

\subsection{Discovery of HC$_7$N$^+$}\label{sec:results_hc7n+}

Given the significant increase in the dipole moment of the HC$_n$H$^+$ series 
with the number of carbon atoms \citep{Zhang2012}, we could expect to detect HC$_7$N$^+$, the following member of the series with $n$ being odd. This molecule has been observed in the optical with a high spectral resolution, allowing the rotational structure of its  $X$$^2\Pi$$\rightarrow$$A$$^2\Pi$ transition to be resolved  \citep{Sinclair2000}. These authors derived $B_0$\,=\,568.6$\pm$0.2 MHz and $A_{SO}$\,=\,$-$36 cm$^{-1}$, which provides an estimate for the effective rotational constant of the $^2\Pi_{3/2}$ ladder of 568.3 MHz.

We have explored the QUIJOTE data around the expected position of the lines of HC$_7$N$^+$ and found eight lines in harmonic relation with half integer quantum numbers from $J_u$=55/2 to 69/2. The lines are shown in the right column of
Fig. \ref{fig_hc5n+} and their line parameters are given in Table \ref{line_parameters_hc7n+}. The lines appear slightly broadened but without any obvious hyperfine splitting, as it could be expected for their high rotational quantum numbers. Moreover, for similarity with HC$_5$N$^+$, the expected $\Lambda$-doubling splitting should be very small.
We have checked that a close shell molecule with $B$\,=\,$B_{\rm eff}$/2 and, hence, that integer quantum numbers cannot
be assigned to the QUIJOTE data. If this were the case, all lines with $J$ even would be missing in our data.

The eight observed lines can be fitted with $B_{\rm eff}$\,=\,567.85036\,$\pm$\,0.00037 MHz and $D_{\rm eff}$\,=\,4.01\,$\pm$\,0.19 Hz (hereafter, referred to as B568). The standard deviation of the
fit is 9.3 kHz. These constants are very close to those of HC$_7$N ($B$\,=\,564.001 MHz
and $D$\,=\,4.04 Hz, \citealt{BizzocchiD2004}) and those of C$_7$N$^-$ \citep[$B$\,=\,582.685 MHz and
$D$\,=\,4.0 Hz,][]{Cernicharo2023a}. The agreement with the optical observations is excellent.
Consequently, we conclude that the
series of lines (B568) pertain to the cation radical HC$_7$N$^+$. Other possible candidates, in particular
the radical C$_7$N, are
discussed and discarded in Appendix\,\ref{candidates_hc7n+}. We have performed quantum chemical calculations for HC$_7$N$^+$ and derived a dipole moment of 7.9\,D (see Appendix\,\ref{ab_initio}).

\subsection{Column densities}\label{column_densities}

Using the derived molecular constants, and for HC$_5$N$^+$ adopting the dipole moment of 6.5\,D derived in this
work (see Appendix\,\ref{ab_initio}),
we derived a rotational temperature, $T_{\rm rot}$, of 5.5\,$\pm$\,0.5\,K and a column density of (9.9\,$\pm$\,1.0)\,$\times$\,10$^{10}$ cm$^{-2}$.
In order to compare the column density of HC$_5$N$^+$ with those of C$_5$N and C$_5$N$^-$,
we have analysed their lines in the QUIJOTE data. These two species were previously
studied by \citet{Cernicharo2020b} with less sensitive data. The observed lines of C$_5$N with
the present version of QUIJOTE
are shown in Fig. \ref{fig_C5N} and those of C$_5$N$^-$ are shown in Fig. \ref{fig_C5N-}
(see Appendix\,\ref{app:C5N_C5N-}).
The line parameters for these two species are given in Table \ref{line_parameters}.
The best
fit to the observed line intensities of C$_5$N has been obtained for
a rotational temperature of 8.6\,$\pm$\,0.2\,K and a column density of (4.8\,$\pm$\,0.3)\,$\times$\,10$^{11}$
cm$^{-2}$. For C$_5$N$^-$ we derived $T_{\rm rot}$\,=\,7.9\,$\pm$\,0.4\,K and $N$\,=\,(8.5\,$\pm$\,0.5)\,$\times$\,10$^{10}$
cm$^{-2}$. The rotational temperatures for these two species have been obtained through a rotational
diagram (see Appendix\,\ref{app:C5N_C5N-}). The value of the column density of C$_5$N$^-$ is a factor of two lower than that derived
by \citet{Cernicharo2020b} due to the different rotational temperatures used for its
determination, with the value previously reported being based on data with a significantly
lower signal-to-noise ratio ($S/N$). The C$_5$N$^-$/HC$_5$N$^+$ abundance ratio is 1.2\,$\pm$\,0.2 and the C$_5$N/HC$_5$N$^+$ one is
4.8\,$\pm$\,0.8.

Finally, a molecule to consider in this analysis is HC$_5$N. A detailed study using
the weak hyperfine satellite lines of all transitions of cyanodiacetylene in the QUIJOTE line survey provides
a rotational temperature of 7.5\,$\pm$\,0.3\,K
and a column density of (6.6\,$\pm$\,0.1)\,$\times$\,10$^{13}$ cm$^{-2}$ (Cernicharo
et al. 2024, in prep.). Hence, the HC$_5$N/HC$_5$N$^+$ abundance
ratio is 670\,$\pm$\,80.

For HC$_7$N$^+$ we have adopted a dipole moment of 7.9\,D (see Appendix\,\ref{ab_initio}) and derived
$T_{\rm rot}$\,=\,8.5\,$\pm$\,0.5\,K and $N$\,=\,(2.3\,$\pm$\,0.2)\,$\times$\,10$^{10}$ cm$^{-2}$.
The column density of HC$_7$N can be derived from all lines of this species within the QUIJOTE data.
We derived $T_{\rm rot}$\,=\,7.8\,$\pm$\,0.1\,K and $N$\,=\,(2.3\,$\pm$\,0.1)\,$\times$\,10$^{13}$ cm$^{-2}$ (Cernicharo et al. 2024, in prep.). Hence, the abundance ratio between HC$_7$N and HC$_7$N$^+$ is 1000\,$\pm$\,150.

The column density of the anion C$_7$N$^-$ is (5.0\,$\pm$\,0.5)\,$\times$\,10$^{10}$ cm$^{-2}$ \citep{Cernicharo2023a},
hence the abundance ratio between C$_7$N$^-$ and HC$_7$N$^+$ is 2.2\,$\pm$\,0.2, that is to say it is similar to that
of C$_5$N$^-$ over HC$_5$N$^+$. Finally, the abundance ratio between HC$_5$N$^+$ and HC$_7$N$^+$
is 4.2\,$\pm$\,0.5, which is of the order of the abundance ratio between the neutral species (2.9\,$\pm$\,0.2).
C$_7$N has not been detected yet (see Appendix\,\ref{candidates_hc7n+}), hence an abundance ratio between this neutral radical and its protonated species cannot be determined.

In addition to HC$_5$N$^+$ and HC$_7$N$^+$, other cyanopolyyne cations could be present in our data, in particular HC$_3$N$^+$. No high spectral resolution optical observations
have been reported for this species in spite of its potential interest in the chemistry, not only of the interstellar medium (ISM),
but also of Titan \citep{Vuitton2006}. The molecule will also have an inverted 
$^2\Pi$ ground electronic state with $A_{SO}$\,=\,$-$44\,$\pm$\,2 cm$^{-1}$ \citep{Gans2016}. Its
rotational constant has been estimated to be between 4565 and 4594 MHz,
depending of the level of theory used in the calculations \citep{Zhang2012}. The predicted
dipole moment is $\sim$5\,D. Only three transitions, $J$=7/2-5/2, $J$=9/2-7/2, and $J$=11/2-9/2, will be within the
frequency coverage of QUIJOTE. These low-$J$ transitions will exhibit a significant hyperfine structure that, although helping in their identification,
will dilute the corresponding line intensities. We could expect, similar to the case of
HC$_5$N$^+$ and HC$_7$N$^+$,
to have a column density $\sim$1/1000 that of the  neutral HC$_3$N. Hence, adopting the column density
derived for HC$_3$N of 1.9$\times$10$^{14}$ cm$^{-2}$ \citep{Tercero2024}, the column density of the cation HC$_3$N$^+$
could be $\sim$1.9$\times$10$^{11}$ cm$^{-2}$. 
Adopting the collisional rates of HC$_3$N as a proxy for the unsplitted levels of HC$_3$N$^+$, then
the expected intensities for the unsplitted transitions
will be  $\sim$40-60 mK (under a LVG calculation). Even when splitted, these lines will be prominent in our data and easily identified.
We have analysed the QUIJOTE data for unidentified lines with frequencies in harmonic relation for a $B_{\rm eff}$ between 4545 and 4615 MHz without
success. Hence,
we conclude that under the adopted assumptions, the abundance of HC$_3$N$^+$ has to be $\le$10$^{11}$ cm$^{-2}$ if its effective rotational constant is within the explored range.

\section{Discussion}\label{sec:discussion}

The cations HC$_5$N$^+$ and HC$_7$N$^+$ can be seen as the protonated forms of C$_5$N and C$_7$N, respectively. These two radicals have a high proton affinity, 870 kJ mol$^{-1}$ and 933.8 kJ mol$^{-1}$, respectively, and thus reactions of proton transfer from abundant cations such as HCO$^+$ are exothermic and can provide a formation pathway to HC$_5$N$^+$ and HC$_7$N$^+$ in TMC-1. Protonation of a neutral species M through efficient proton donors such as HCO$^+$ results in abundance ratios MH$^+$/M in the range 10$^{-3}$-10$^{-1}$ \citep{Agundez2022}, while the abundance ratio HC$_5$N$^+$/C$_5$N of $\sim$\,0.2 derived in TMC-1 is higher, which probably implies that proton transfer is not the main formation route to HC$_5$N$^+$. There is however one important aspect to consider that is related to the dipole moment of C$_5$N. As discussed previously \citep{Cernicharo2008,Cernicharo2020b}, the true dipole moment of C$_5$N could be an average between that of its ground state $^2\Sigma$, which has been calculated to be 3.385\,D \citep{Botschwina1996}, and that of its first electronic excited state $^2\Pi$, which lies very low in energy ($\sim$500 cm$^{-1}$; \citealt{Botschwina1996}) and has a dipole moment of $\sim$0.2-1\,D \citep{Pauzat1991,Cernicharo2008}. Hence, C$_5$N could have a dipole moment between these two values in the case of admixing between the $^2\Sigma$ and the $^2\Pi$ states. If so, the column density of C$_5$N would increase by a factor of four and the HC$_5$N$^+$/C$_5$N ratio would decrease to $\sim$\,0.05. In this case the C$_5$N$^-$/C$_5$N abundance ratio would also decrease by a factor of four, from 0.43 to 0.1, in better agreement with anion-to-neutral ratios of species of a similar size \citep{Agundez2023b}. A similar effect has been analysed for C$_4$H by \citet{Oyama2020}, leading to a revised value of its dipole moment that removes previous inconsistencies in the column densities derived for this molecule in different astrophysical environments.

\begin{figure}
\centering
\includegraphics[width=\columnwidth]{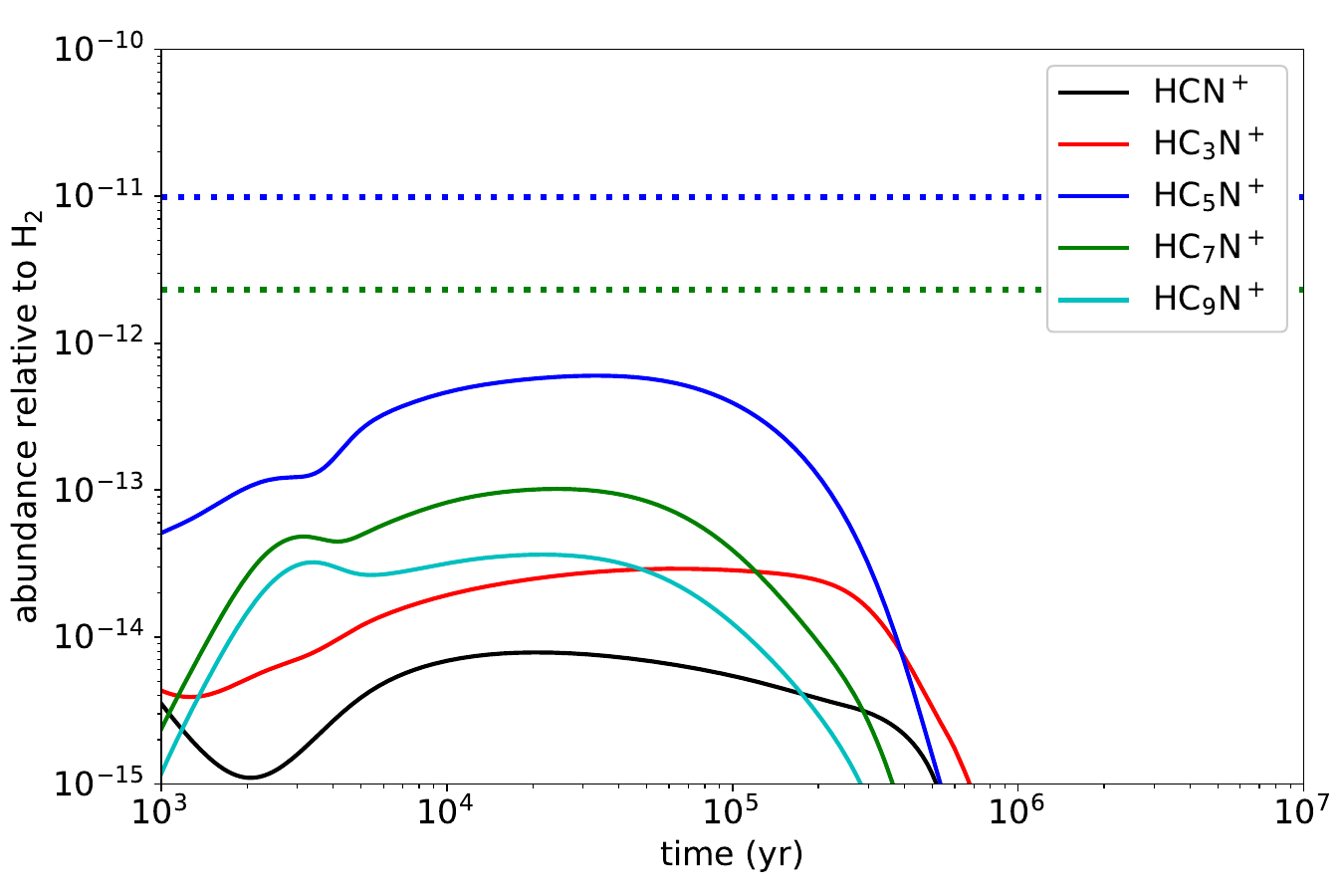}
\caption{Calculated fractional abundances for the series of ionized cyanopolyynes HCN$^+$, HC$_3$N$^+$, HC$_5$N$^+$, HC$_7$N$^+$, and HC$_9$N$^+$ as a function of time. Horizontal dotted lines correspond to the values observed in TMC-1 for HC$_5$N$^+$ and HC$_7$N$^+$, adopting a column density of H$_2$ of 10$^{22}$ cm$^{-2}$ \citep{Cernicharo1987}.}
\label{fig:abun}
\end{figure}

In order to shed additional light on the formation of HC$_5$N$^+$ and HC$_7$N$^+$, we have carried 
out chemical modelling calculations using typical parameters of cold dense clouds \citep{Agundez2013} 
and the chemical network from the latest release of the UMIST database for astrochemistry \citep{Millar2024}. 
According to the chemical model, the most important reactions in the formation of cyanopolyynes 
cations HC$_n$N$^+$ ($n$\,=\,1, 3, 5, 7, 9, ...) are 
\begin{equation}
\rm H_2 + C_{\textit n}N^+ \rightarrow HC_{\textit n}N^+ + H,
\end{equation}
\begin{equation}
\rm H^+ + HC_{\textit n}N \rightarrow HC_{\textit n}N^+ + H,
\end{equation}
while the reactions of proton transfer from HCO$^+$, H$_3$O$^+$, and H$_3^+$ to the radicals C$_n$N contribute as well, 
but to a lower extent. The calculated abundances for the series of cations HC$_n$N$^+$ are shown as a function of 
time in Fig.\,\ref{fig:abun}, where we also plotted the abundances of HC$_5$N$^+$ and HC$_7$N$^+$ observed 
in TMC-1. It is seen that the cations HC$_n$N$^+$ appear early, after a few 10$^3$ yr, and 
last until approximately 10$^5$ yr. However, the peak abundances calculated for HC$_5$N$^+$ and 
HC$_7$N$^+$ lie about 20 times below the observed values.

We have explored the possibility that additional formation routes are missing in the UMIST network. 
For instance, the reactions of CN with the cations C$_4$H$_2^+$ and C$_6$H$_2^+$ could be a 
source of HC$_5$N$^+$ and HC$_7$N$^+$, respectively. These reactions are not included in the 
UMIST database and to our knowledge have not been studied. However, the analogous reaction 
CN + C$_2$H$_2$$^+$ is included in the UMIST network with a high rate coefficient, estimated 
originally by \cite{Prasad1980}, and theoretical calculations support that it is fast at low 
temperatures and yields HC$_3$N$^+$ as a main product \citep{Redondo2000}. If we include these 
reactions, they contribute somewhat to the formation of HC$_5$N$^+$ and HC$_7$N$^+$, although 
the calculated abundances of these cations do not experience any appreciable enhancement.

The main destruction path of the cations HC$_n$N$^+$ at times between 10$^4$ and 10$^6$ yr is 
the reaction with H$_2$, and to a lower extent (10-50 times less) the dissociative recombination 
with electrons. There are only a few molecules detected in the ISM that react with H$_2$. 
A good example is C$_3$H$^+$ \citep{Pety2012,Cernicharo2022a}, which is known to react with H$_2$ 
with a moderate rate coefficient of 2.6\,$\times$\,10$^{-11}$ cm$^3$ s$^{-1}$ \citep{McEwan1999}. 
The larger analogous species C$_5$H$^+$ is assumed to react very slowly with H$_2$. The cation 
HC$_3$N$^+$ is known to react with H$_2$ with a relatively low rate coefficient of 
7\,$\times$\,10$^{-12}$ cm$^3$ s$^{-1}$ \citep{Knight1985}, and longer analogous cations 
HC$_n$N$^+$ are assumed to react with H$_2$ with a similar rate coefficient of 5\,$\times$\,10$^{-12}$ 
cm$^3$ s$^{-1}$ (the rate coefficient of the reaction HC$_5$N$^+$ + H$_2$, taken as 
10$^{-9}$ cm$^3$ s$^{-1}$ in the UMIST network, was decreased to 5\,$\times$\,10$^{-12}$ cm$^3$ s$^{-1}$). 
It could happen that the underestimation of the abundances of HC$_5$N$^+$ and HC$_7$N$^+$ in the chemical 
model is not due to missing formation routes but to a too fast destruction through their reaction with H$_2$. 
It would therefore be interesting to investigate the rate coefficient of 
the reactions of HC$_5$N$^+$ and HC$_7$N$^+$ (and also C$_5$H$^+$) with H$_2$.

\begin{acknowledgements}
We thank Ministerio de Ciencia e Innovaci\'on of Spain (MICIU) for funding support through projects 
PID2019-106110GB-I00, and PID2019-106235GB-I00. We also thank ERC for funding through grant 
ERC-2013-Syg-610256-NANOCOSMOS. We thank the Consejo Superior de Investigaciones Cient\'ificas 
(CSIC; Spain) for funding through project PIE 202250I097. The present study was also supported by 
Ministry of Science and Technology of Taiwan and Consejo Superior de Investigaciones Cient\'ificas 
under the MoST-CSIC Mobility Action 2021 (Grants 11-2927-I-A49-502 and OSTW200006).

\end{acknowledgements}

\textbf{Note added to proofs}
We have recently found a series of lines that we have been able to
assign to HC$_3$N$^+$. The lines appear below the frequency ranges explored
in this work. The result will be submitted soon (Cabezas et al. 2024, in prep.),
and confirms the presence of cationic cyanopolyynes in TMC-1.

\normalsize

\begin{appendix}

\section{Line parameters}\label{app:lineparameters}

Line parameters for all observed transitions with the Yebes 40m radio telescope
have been derived by fitting a Gaussian line profile to them
using the GILDAS package. A
velocity range of $\pm$20\,\kms\, around each feature was considered for the fit after a polynomial
baseline was removed. Negative features produced in the folding of the frequency switching data were blanked
before baseline removal.

The derived frequency centroids of the lines of HC$_5$N$^+$ are given, together with their
integrated intensities and equivalent line widths, in Table \ref{line_centroid}. The derived line
parameters for HC$_7$N$^+$ are given in Table \ref{line_parameters_hc7n+}.

\begin{table}
\tiny
\centering
\caption{Estimated frequency centroid of the observed lines of HC$_5$N$^+$.}
\label{line_centroid}
\begin{tabular}{lcccc}
\hline
\hline
$J_u-J_l$    & $\nu_{obs}$~$^a$    & $\int T_A^* dv$~$^b$ & <v$_{LSR}$>$^c$ & $\Delta$v$^d$\\
             &      (MHz)          & (mK km\,s$^{-1}$)    &         &              \\
\hline
    25/2-23/2& 33416.338$\pm$0.040 & 4.3$\pm$0.4          & 5.8    & 2.8\\
    27/2-25/2& 36089.623$\pm$0.040 & 4.4$\pm$0.4          & 5.8    & 2.3\\
    29/2-27/2& 38762.821$\pm$0.040 & 3.7$\pm$0.4          & 5.9    & 1.9\\
    31/2-29/2& 41436.111$\pm$0.040 & 3.0$\pm$0.3          & 5.8    & 1.7\\
    33/2-31/2& 44109.381$\pm$0.040 & 1.7$\pm$0.2          & 5.6    & 1.0\\
    35/2-33/2& 46782.597$\pm$0.040 & 2.2$\pm$0.2          & 5.7    & 1.4\\
    37/2-35/2& 49455.783$\pm$0.040 & 0.6$\pm$0.1          & 5.8    & 0.7\\
\hline
\end{tabular}
\tablefoot{
\tablefoottext{a}{Frequency centroid of the line. We have adopted a
v$_{LSR}$ of 5.83 km\,s$^{-1}$ for the source \citep{Cernicharo2020c}.}
\tablefoottext{b}{Integrated line intensity in mK\,km\,s$^{-1}$.}
\tablefoottext{c}{Averaged velocity (in km~s$^{-1}$).}
\tablefoottext{d}{Equivalent line width (in km~s$^{-1}$).}
}
\end{table}
\normalsize
\begin{table}
\tiny
\centering
\caption{Observed line parameters of HC$_7$N$^+$.}
\label{line_parameters_hc7n+}
\begin{tabular}{lccccc}
\hline
\hline
$J_u-J_l$    & $\nu_{obs}$~$^a$    & (o-c)$^b$ & $\int T_A^* dv$~$^c$ &  $\Delta$v$^d$& T$_A$$^*$$^e$\\
             &      (MHz)          &  (kHz)    & (mK km\,s$^{-1}$)    &               &  (mK)    \\
\hline
55/2-53/2& 31231.441              &  5.0 &  0.90$\pm$0.12  &0.79$\pm$0.12 &  1.07$\pm$0.14 \\
57/2-55/2& 32367.105              &  6.0 &  0.76$\pm$0.12  &0.83$\pm$0.15 &  0.87$\pm$0.07 \\
59/2-57/2& 33502.745              &-14.2 &  0.59$\pm$0.06  &0.86$\pm$0.09 &  0.65$\pm$0.05 \\
61/2-59/2& 34638.423              &  6.4 &  0.43$\pm$0.04  &0.96$\pm$0.10 &  0.41$\pm$0.07 \\
63/2-61/2& 35774.073              &  2.0 &  0.40$\pm$0.12  &0.65$\pm$0.13 &  0.57$\pm$0.08 \\
65/2-63/2& 36909.712              &-10.5 &  0.56$\pm$0.03  &0.72$\pm$0.05 &  0.73$\pm$0.05 \\
67/2-65/2& 38045.367              & -3.8 &  0.44$\pm$0.10  &0.93$\pm$0.25 &  0.45$\pm$0.08 \\
69/2-67/2& 39181.025              &  9.2 &  0.21$\pm$0.05  &0.46$\pm$0.19 &  0.43$\pm$0.09 \\
\hline
\end{tabular}
\tablefoot{
\tablefoottext{a}{Measured frequency of the line. The uncertitude is 10 kHz for
all lines. We have adopted a
v$_{LSR}$ of 5.83 km\,s$^{-1}$ for the source \citep{Cernicharo2020c}.}
\tablefoottext{b}{Observed minus calculated frequencies (in kHz).}
\tablefoottext{c}{Integrated line intensity in mK\,km\,s$^{-1}$.}
\tablefoottext{d}{Line width (in km~s$^{-1}$).}
\tablefoottext{e}{Antenna temperature (mK).}
}
\end{table}
\normalsize

\section{Possible candidates for B1336}\label{candidates_hc5n+}
Although the agreement between our rotational constant for B1336 with that derived from
the optical observations of \citet{Sinclair1999a} is excellent, it is important to discard
other possible candidates that could fit the rotational constant and be abundant enough in
TMC-1.
The isotopologues of HC$_5$N have rotational constants below that of the main isotopologue
and have integer quantum numbers. Moreover, no vibrationally bending excited states of HC$_5$N, which have rotational constants larger than that of the ground state, could be responsible for
the observed lines as these states will also have integer quantum numbers.
Another molecule with close rotational and distortion constants is
the C$_5$N$^-$ anion ($B$=1388.867 MHz, $D$=35.7 Hz; \citealt{Cernicharo2008,Cernicharo2020b}).
However, this species is a closed shell molecule. An interesting species to consider is
the radical C$_5$N, which also has close rotational and distortion constants ($B$=1403.08 MHz,
$D$=50$\pm$10 Hz; \citealt{Kasai1997}) and has been detected in TMC-1 \citep{Guelin1998}. Nevertheless, this radical has a $^2\Sigma$
ground state with integer values for $N$. Some of the bending vibrationally excited states of C$_5$N will have a $^2\Pi$ vibronic character with half-integer quantum numbers. Such
excited states have been found in TMC-1 for C$_6$H in its
low-energy $\nu_{11}$ bending mode \citep{Cernicharo2023b}. However, similarly to C$_6$H, we expect that
the vibrationally excited states of C$_5$N will have rotational constants larger than that of
the ground state. Moreover, C$_5$N is less abundant than C$_6$H. Hence, the possibility that our lines are produced by 
C$_5$N in a bending state can be disregarded.
Other possible candidates with a
rotational constant close to our value are C$_6$H$^+$ and C$_5$N$^+$. However, their expected ground electronic state is a bent $^3A''$ with (B+C)/2$\sim$1392.6 MHz and 1394.7 MHz, respectively \citep{Aoki2014}.
These two molecules will produce lines with integer quantum numbers and can be 
discarded.

Some species
such as NC$_4$N$^+$ and HC$_6$H$^+$ have $^2\Pi$ inverted ground electronic states,
with $B$=1339.7$\pm$0.4 MHz \citep{Sinclair1999a} and $B$=1336.9$\pm$0.1 MHz \citep{Sinclair1999b}, respectively.
Unfortunately, these molecular species are symmetric and lack a permanent dipole moment. Hence, in
spite of the good match in the rotational constant, they cannot be the carrier of our lines. HC$_6$D$^+$ will
have a small dipole moment. However, its rotational constant will be too low and the required column density
too large to explain our lines. Another species with similar rotational and distortion constants
is NC$_4$NH$^+$ \citep[$B$=1293.9 MHz, $D$=29.8 Hz;][]{Agundez2023a}. Although this species is a
close shell molecule, it gives, together with those discussed above, a good indication
of the structure and composition of the carrier of our lines. Species containing sulphur such as
HCCCCS, which has been found in TMC-1 \citep{Fuentetaja2023}, can be discarded as its rotational constant is too large, $B$=1435.3 MHz \citep{Hirahara1994}. Its lines are too weak to permit the detection of its
deuterated isotopologue which could have a rotational constant around our $B$. Carbon chains containing
oxygen such as HC$_4$O \citep[$B\sim$2245 MHz;][]{Kohguchi1994} or
HC$_5$O \citep[$B\sim$1293.6 MHz, $D$=36 Hz;][]{Mohamed2005}, 
can also be discarded. HC$_5$O has been been detected in TMC-1 by \citet{McGuire2017}.

Taking into account the observational facts and the arguments developed above, we conclude that the most likely carrier has to be very close in structure to HC$_5$N (which has a very
close rotational constant) and C$_5$N. The most plausible candidate is the radical HC$_5$N$^+$, the protonated species of C$_5$N.

\section{Molecular parameters for HC$_5$N$^+$ and HC$_7$N$^+$ from quantum chemical calculations}\label{ab_initio}

Several theoretical calculations for HC$_5$N$^+$ and HC$_7$N$^+$ species have been previously reported. However, we carried out different quantum chemical calculations for HC$_5$N$^+$ and HC$_7$N$^+$ molecules in order to estimate some spectroscopic parameters not reported in the literature before. We optimized the geometry of the HC$_5$N$^+$ radical at
the same level of theory employed by \citet{Gans2019}, ic-MRCI+Q/AVTZ, using the MOLPRO 2020 ab initio program package \citep{Werner2020}. The values for the $B_e$ rotational constant and the electric dipole moment of HC$_5$N$^+$ using this level of theory are 1339.7\,MHz and 6.5\,D, respectively. These values slightly differ from those reported by \citet{Gans2019}. At the optimized geometry, we calculated the values for the hyperfine coupling constants at the B3LYP/cc-pVTZ level of theory using the GAUSSIAN 16 package \citep{Frisch2016}. 
The Frosch and Foley \citep{Frosch1952} hyperfine constants were derived from the dipole-dipole interaction tensor 
\textbf{T} and the Fermi contact constant $b_F$ using the relations as

\begin{eqnarray}
   b = {b_F - {c\over 3}} = {b_F - {1\over 2} T_{aa}}
\end{eqnarray}
\begin{eqnarray}
   c = {{3\over 2} T_{aa}}
\end{eqnarray}
\begin{eqnarray}
   d = { T_{bb} - T_{cc}}
.\end{eqnarray}

We utilized the approximate relation \citep{Townes1975}

\begin{eqnarray}
  c = 3(a-d),
\end{eqnarray}
\noindent
between $a$, $c$, and $d$ to estimate the $a$ constant as

\begin{eqnarray}
   a = {{c\over 3} + d} = {{3\over 2} T_{aa} + T_{bb}}
.\end{eqnarray}

The main contribution to the hyperfine coupling in the $^2\Pi_{3/2}$ state comes from the $h_1$ constant, defined as

\begin{eqnarray}
h_1 = {a + {(b + c) \over 2}}
.\end{eqnarray}

\begin{figure}[h]
\centering
\includegraphics[width=0.48\textwidth]{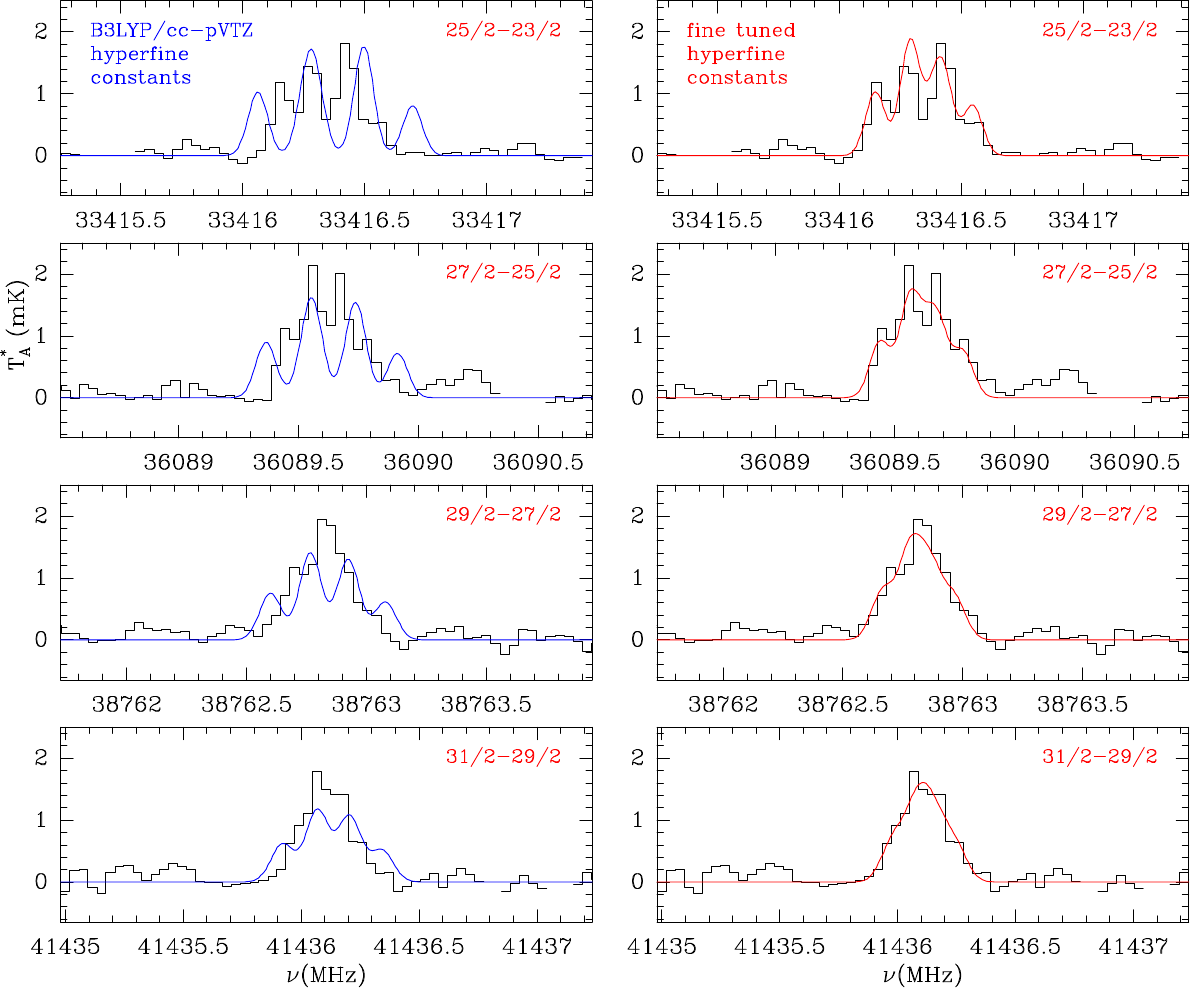}
\caption{Same as Fig. \ref{fig_hc5n+}, but comparing observations with modelled spectra using the
hyperfine constants of HC$_5$N$^+$ derived from our quantum chemical calculations (see App. \ref{ab_initio}).
The blue spectra in the left panels show the modelled lines. The panels of the right
column show in red the final modelled spectra after fine-tuning the hyperfine constants (identical
to those shown in Fig. \ref{fig_hc5n+}).
The centroids of the modelled lines of HC$_5$N$^+$ in both columns have been fixed to the observed ones (see Table \ref{line_centroid}).}
\label{simul}
\end{figure}

The hyperfine coupling constants obtained for HC$_5$N$^+$ are shown in Table \ref{hyperfine}. They were 
employed to predict the expected hyperfine pattern of HC$_5$N$^+$, as shown in the left panels of 
Figure \ref{simul}. As it can be seen, the predicted hyperfine splittings are larger than those observed, 
which means that the $h_1$ constant for H and N are larger than those predicted. To obtain a better 
reproduction of the observed splitting of HC$_5$N$^+$, we systematically decreased the $h_1$ constants 
for H and N with the same factor. The best agreement between the simulation and the observations are 
obtained with $h_1$ values 1.6 times smaller than those predicted by our calculations. The hyperfine 
coupling constants needed to reproduce the observations are shown in Table \ref{hyperfine}. As it can 
be seen the b$_F$, T$_{aa}$ and T$_{bb}$ modified constants are not very different to those predicted 
by our calculations, which is additional proof supporting the identification of HC$_5$N$^+$ as the carrier 
of our observed lines.

In the case of HC$_7$N$^+$, we carried out molecular structure optimization calculations at the RCCSD(T)/cc-pVTZ level 
of theory using the MOLPRO 2020 ab initio program package \citep{Werner2020}. The calculated values for the rotational 
constant $B_e$ and the electric dipole moment are 561.8 MHz and 7.9 D, respectively. As in the case of HC$_5$N$^+$, 
we also estimated the values of the  hyperfine coupling constants at the B3LYP/cc-pVTZ level of theory using the 
GAUSSIAN 16 package \citep{Frisch2016}. Our predictions using those constants show that the hyperfine structures are
completely collapsed for all the expected lines of HC$_7$N$^*$ in the Q band, in agreement with our observations.

\begin{table}
\small
\centering
\caption{Hyperfine coupling constants of HC$_5$N$^+$ (all in MHz).}
\label{hyperfine}
\centering
\begin{tabular}{{lcc}}
\hline
\hline
Constant              & B3LYP/cc-pVTZ & Best simulation \\
\hline
$b_F$(H)              &  -30.98    &     -33.04   \\
$T_{aa}$(H)           &  18.44     &     17.42    \\
$T_{bb}$(H)           &  -2.80     &     -4.03    \\
$a$(H)                &  22.07     &     18.07    \\
$b$(H)                & -40.20     &     -41.75   \\
$c$(H)                &  27.67     &     26.13    \\
$h_1$(H)              &  15.81     &     10.26    \\
\hline
$b_F$(N)              &  0.57      &     -2.03    \\
$T_{aa}$(N)           &  -15.06    &     -17.00   \\
$T_{bb}$(N)           &  26.65     &     24.31    \\
$a$(N)                &  30.71     &     23.12    \\
$b$(N)                &  8.09      &      6.47    \\
$c$(N)                & -22.59     &     -25.50   \\
$h_1$(N)                &  23.46     &     13.61    \\
\hline
\hline
\end{tabular}
\end{table}
\normalsize

\section{Possible candidates for B568}\label{candidates_hc7n+}
Similar to HC$_5$N$^+$, several species with a $^2\Pi$ electronic ground state could have rotational constants close to that of B568. As commented on in Sect. \ref{sec:results_hc7n+}, the rotational
constant of B568 is very close to that of the closed shell species HC$_7$N and C$_7$N$^-$, which suggests that B568 is
structurally very similar to these species. Isotopologues or vibrationally excited
states of these species have integer quantum numbers and can be discarded.

An important species that should be present, although no detected yet, is C$_7$N. Its
anionic form, C$_7$N$^-$, has been detected in TMC-1 and IRC+10216 \citep{Cernicharo2023a}. Ab initio
calculations by \citet{Botschwina1997} show a peculiar effect for this species, with a ground
electronic state $^2\Pi$ having a low dipole moment ($\sim$1\,D) and an electronic excited state $^2\Sigma$ at
$\sim$300 cm$^{-1}$ above, and with a large dipole moment. The rotational
constants for the $^2\Pi$ and $^2\Sigma$ states are $\sim$585 and $\sim$583.1 MHz, respectively. The estimated
uncertainty on these constants is 2 MHz \citep{Botschwina1997}.
It was not clear from these calculations what electronic ground state could have C$_7$N.
Even if C$_7$N has a $^2\Pi$ state, it does not fit B568 as
the difference between the predicted and observed rotational constants is $\sim$20 MHz.
Moreover, if the ground state of C$_7$N is $^2\Pi$, then we could expect a significant $\Lambda$-doubling
splitting in its $^2\Pi_{1/2}$ and $^2\Pi_{3/2}$ ladders due to the presence of the very close $^2\Sigma$ state.
Hence, C$_7$N cannot be the carrier of B568.

We have explored our data searching for C$_7$N but
not obvious lines can be assigned at the present level of sensitivity. Due to
the low dipole moment of the $^2\Pi$ state, its lines will be well below our 3$\sigma$ detection limit.
We have also searched for a series of doublets with $B$=580-590 MHz that could arise from the $^2\Sigma$
state without success.

HC$_7$O has a $^2\Pi$ ground electronic state, but its
rotational constant has been measured in the laboratory to be 549.2 MHz \citep{Mohamed2005}. This molecule
has been detected in TMC-1 \citep{Cordiner2017,Cernicharo2021d}, but its isotopologues will have rotational 
constants that are too low. Finally, considering S-bearing species, the first molecule with a rotational constant 
close to that of
B568 is the HC$_6$S $^2\Pi$ radical. 
It has been observed in the laboratory and its rotational constant is of 572.1 MHz \citep{Gordon2002}.
The same authors also observed H$_2$C$_6$S which has ($B$+$C$)/2=559.8 MHz, but the molecule is a close shell
asymmetric species. None of these two species are detected in our data. Hence, we can exclude their isotopologues 
or vibrationally excited states as possible carriers of B568.
Hence, we have to conclude, as explained in Sect. \ref{sec:results_hc7n+}, 
that the best candidate for the
carrier of the lines of B568 is the cationic radical HC$_7$N$^+$.

\section{C$_5$N and C$_5$N$^-$}\label{app:C5N_C5N-}
\begin{figure}[h]
\centering
\includegraphics[width=0.47\textwidth]{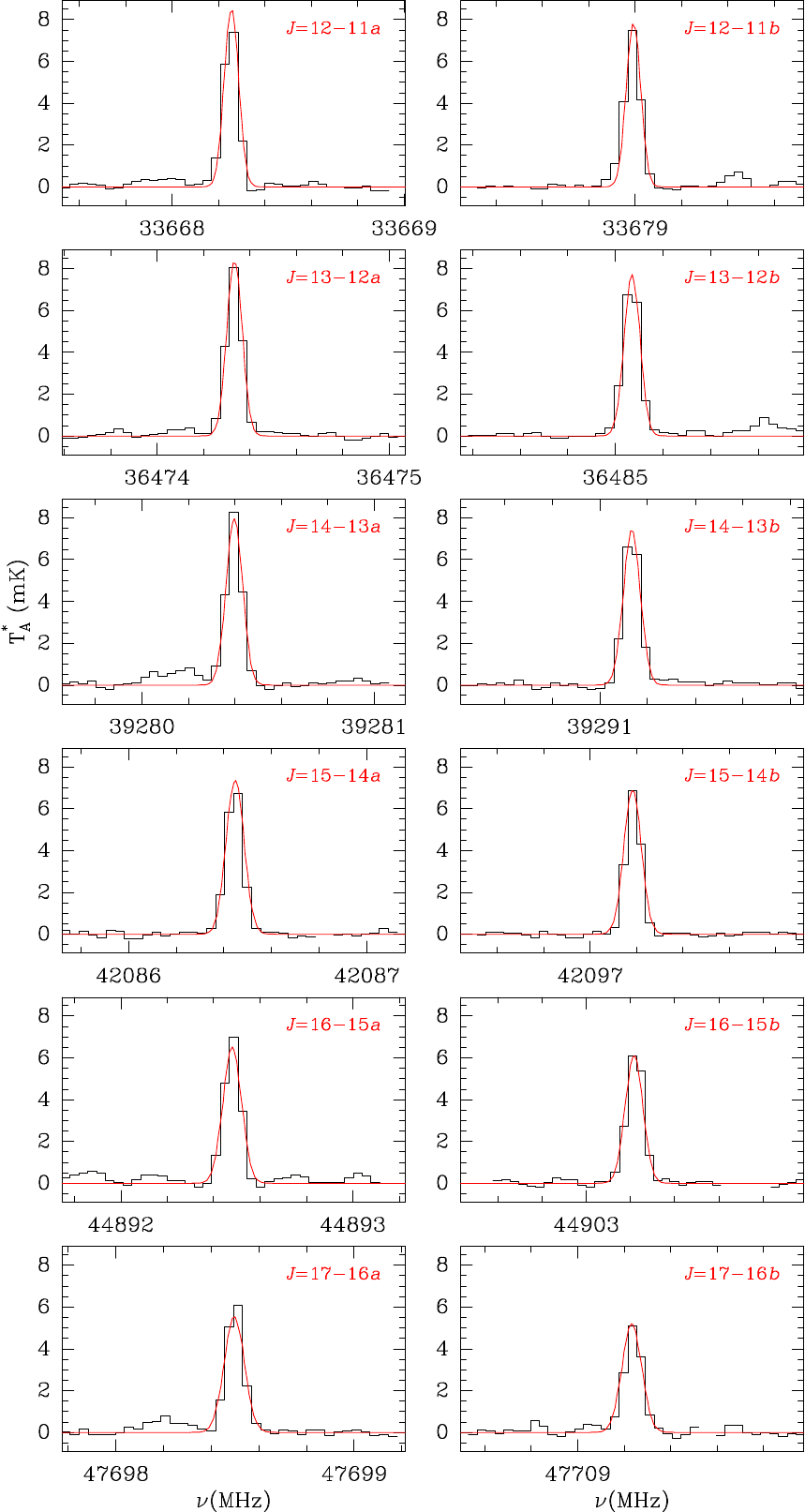}
\caption{Observed transitions of C$_5$N in TMC-1. Quantum numbers are indicated at the
top right of each panel.
The abscissa corresponds to the rest frequency adopting a velocity for the
source of 5.83 km\,s$^{-1}$ \citep{Cernicharo2020c}. The ordinate is the antenna temperature, corrected for
atmospheric and telescope losses, in milli Kelvin.
Blank channels
correspond to negative features produced when folding the frequency-switched data.
The red line shows the modelled spectra for these lines (see App. \ref{app:C5N_C5N-}).}
\label{fig_C5N}
\end{figure}

\begin{figure}[h]
\centering
\includegraphics[width=0.47\textwidth]{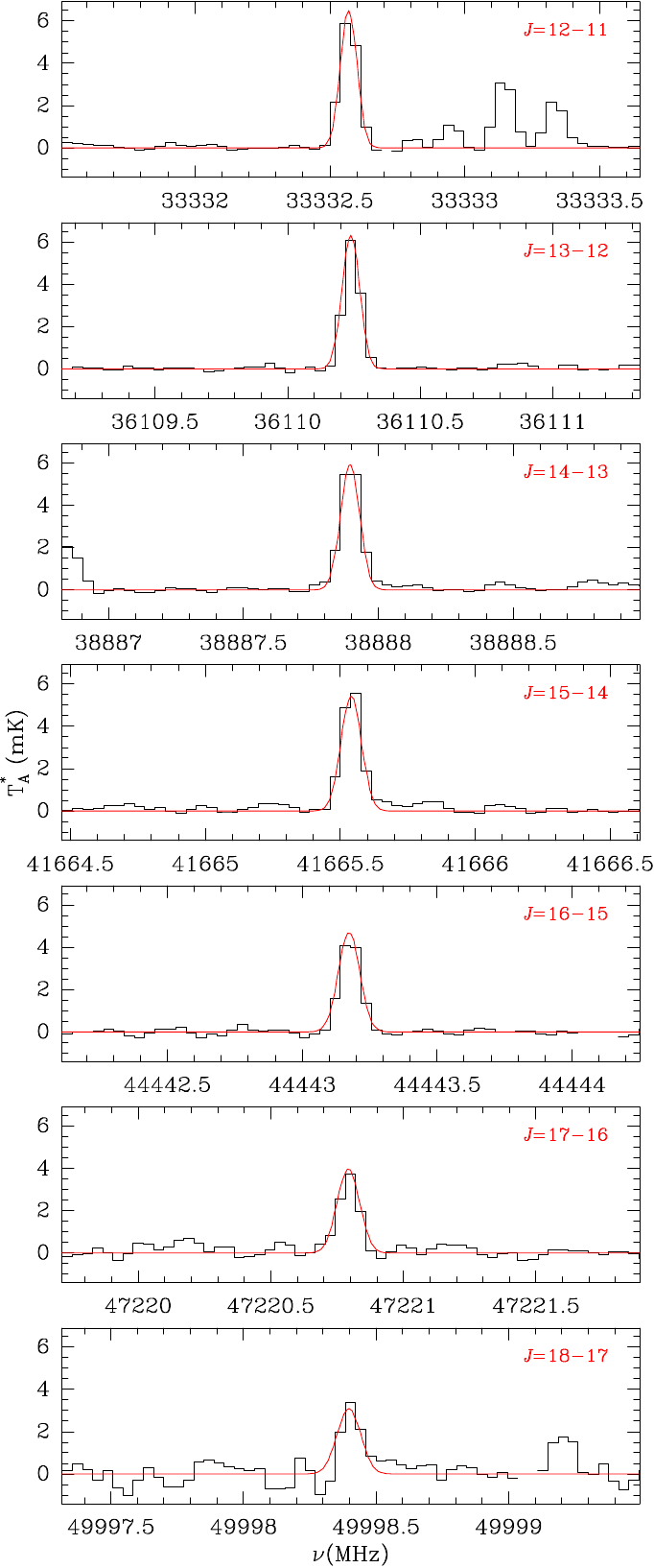}
\caption{Observed transitions of C$_5$N$^-$ in TMC-1. Quantum numbers are indicated at the
top right of each panel.
The abscissa corresponds to the rest frequency adopting a velocity for the
source of 5.83 km\,s$^{-1}$ \citep{Cernicharo2020c}. The ordinate is the antenna temperature, corrected for
atmospheric and telescope losses, in mK.
Blank
channels correspond to negative features produced when folding the frequency-switched data.
The red line shows the modelled spectra for these lines (see App. \ref{app:C5N_C5N-}).}
\label{fig_C5N-}
\end{figure}

In this work we have re-analysed the lines of C$_5$N and C$_5$N$^-$ previously published by \citet{Cernicharo2020b}.
The lines of C$_5$N are shown in Fig. \ref{fig_C5N} and those of C$_5$N$^-$ in Fig. \ref{fig_C5N-}.
The derived line parameters are given in Table \ref{line_parameters}.
We have used a rotational diagram to derive the rotational temperature of these species and found
$T_{\rm rot}$(C$_5$N)=8.6$\pm$0.2\,K and $T_{\rm rot}$(C$_5$N$^-$)=7.9$\pm$0.4\,K.
The best fit to the column densities are N(C$_5$N)=(4.8$\pm$0.3)$\times$10$^{11}$
and N(C$_5$N$^-$)=(8.5$\pm$0.5)$\times$10$^{10}$ cm$^{-2}$.
The value of the column density of C$_5$N$^-$ is a factor two lower than that derived
by \citet{Cernicharo2020b} due to the different rotational temperatures used for its
determination. The value previously reported, $T_{\rm rot}$=6.3$\pm$0.5\,K, was derived from data with a significant lower sensitivity.

The data used in this work for  C$_7$N$^-$ are those of \citet{Cernicharo2023a}. The present
data do not provide a significant improvement to the S/N. 
Hence, the previously derived column density for this species does not
require further revision.

\onecolumn

\begin{table*}[h]
\centering
\caption{Observed line parameters for C$_5$N and C$_5$N$^-$.}
\label{line_parameters}
\begin{tabular}{lcccrccr}
\hline
$J_u-J_l$  & $\nu_{rest}$~$^a$ & $\int T_A^* dv$~$^b$ & v$_{LSR}$$^c$       & $\Delta v$~$^d$ & $T_A^*$~$^e$\\
        & (MHz)              & (mK\,km\,s$^{-1}$)  & (km\,s$^{-1}$)  & (km\,s$^{-1}$)  & (mK) & \\
\hline
C$_5$N\\
12-11a & 33668.256$\pm$0.003 & 5.75$\pm$0.14 & 5.87$\pm$0.01 &    0.68$\pm$0.02&  7.97$\pm$0.09\\
12-11b & 33678.992$\pm$0.003 & 5.91$\pm$0.06 & 5.88$\pm$0.01 &    0.75$\pm$0.01&  7.40$\pm$0.07\\
13-12a & 36474.334$\pm$0.003 & 5.78$\pm$0.06 & 5.86$\pm$0.01 &    0.67$\pm$0.01&  8.06$\pm$0.07\\
13-12b & 36485.070$\pm$0.003 & 5.58$\pm$0.09 & 5.85$\pm$0.01 &    0.69$\pm$0.01&  7.61$\pm$0.08\\
14-13a & 39280.398$\pm$0.003 & 5.39$\pm$0.11 & 5.85$\pm$0.01 &    0.62$\pm$0.01&  8.22$\pm$0.13\\
14-13b & 39291.134$\pm$0.003 & 5.02$\pm$0.11 & 5.84$\pm$0.01 &    0.63$\pm$0.01&  7.46$\pm$0.13\\
15-14a & 42086.447$\pm$0.003 & 4.66$\pm$0.08 & 5.85$\pm$0.01 &    0.60$\pm$0.01&  7.26$\pm$0.12\\
15-14b & 42097.183$\pm$0.003 & 4.24$\pm$0.09 & 5.82$\pm$0.01 &    0.57$\pm$0.01&  6.99$\pm$0.10\\
16-15a & 44892.480$\pm$0.003 & 4.16$\pm$0.09 & 5.83$\pm$0.01 &    0.56$\pm$0.01&  6.99$\pm$0.15\\
16-15b & 44903.216$\pm$0.003 & 4.15$\pm$0.09 & 5.81$\pm$0.01 &    0.59$\pm$0.01&  6.57$\pm$0.11\\
17-16a & 47698.497$\pm$0.003 & 3.69$\pm$0.07 & 5.81$\pm$0.01 &    0.54$\pm$0.01&  6.46$\pm$0.12\\
17-16b & 47709.233$\pm$0.003 & 3.20$\pm$0.11 & 5.80$\pm$0.01 &    0.59$\pm$0.03&  5.14$\pm$0.16\\
\\
C$_5$N$^-$\\
12-11  & 33332.570$\pm$0.004 & 4.85$\pm$0.08 & 5.82$\pm$0.01 &    0.71$\pm$0.01&  6.38$\pm$0.07\\
13-12  & 36110.238$\pm$0.004 & 4.11$\pm$0.06 & 5.79$\pm$0.01 &    0.63$\pm$0.01&  6.15$\pm$0.07\\
14-13  & 38887.896$\pm$0.004 & 4.47$\pm$0.10 & 5.82$\pm$0.01 &    0.68$\pm$0.01&  6.21$\pm$0.09\\
15-14  & 41665.541$\pm$0.004 & 3.73$\pm$0.09 & 5.83$\pm$0.01 &    0.59$\pm$0.01&  5.90$\pm$0.09\\
16-15  & 44443.174$\pm$0.004 & 2.90$\pm$0.09 & 5.80$\pm$0.01 &    0.59$\pm$0.02&  4.62$\pm$0.12\\
17-16  & 47220.793$\pm$0.004 & 2.27$\pm$0.10 & 5.82$\pm$0.01 &    0.57$\pm$0.03&  3.73$\pm$0.16\\
18-17  & 49998.398$\pm$0.005 & 1.89$\pm$0.25 & 5.76$\pm$0.04 &    0.52$\pm$0.08&  3.40$\pm$0.47\\
\hline
\end{tabular}
\tablefoot{
\tablefoottext{a}{Rest frequency.}\\
\tablefoottext{b}{Integrated intensity (in mK km\,s$^{-1}$).}\\
\tablefoottext{c}{v$_{LSR}$ on the line (in km\,s$^{-1}$).}\\
\tablefoottext{d}{Line width (in km\,s$^{-1}$).}\\
\tablefoottext{e}{Antenna temperature (in mK).}
}
\end{table*}

\end{appendix}
\end{document}